\documentclass[aps,superscriptaddress,showpacs,nofootinbib]{revtex4}

\usepackage{epsfig}
\input{epsf}

\begin{document}

\title{Absence of isentropic expansion in various inflation models}

\author{Arjun Berera}
%\email{ab@ph.ed.ac.uk}
\affiliation{ School of Physics, University of
Edinburgh, Edinburgh, EH9 3JZ, United Kingdom}

\author{Rudnei O. Ramos}
%\email{rudnei@uerj.br}
\affiliation{Departamento de F\'{\i}sica Te\'orica,
Universidade do Estado do Rio de Janeiro,
20550-013 Rio de Janeiro, RJ, Brazil}

\begin{abstract}

Dynamics of the inflaton is studied when it interacts with 
boson and fermion fields and in minimal supersymmetric
models. This encompasses multifield inflation models, such as hybrid
inflation, and typical reheating models. {}For much 
of the parameter space
conducive to inflation, the inflaton is found to dissipate adequate
radiation to have observational effects on density perturbations and in
cases to significantly affect inflaton evolution. Thus, for many
inflation models, believed to yield exclusively isentropic inflation,
the parameter space now divides into regimes of 
isentropic and
nonisentropic inflation.

\medskip

\medskip

\noindent
keywords: cosmology, inflation, field dynamics, dissipation
\end{abstract}


\maketitle

\medskip

In Press Physics Letters B 2005

\section{Introduction}

The standard picture of inflation \cite{oldi,ni,ci} has two distinct
regimes. In the first phase, the universe undergoes inflationary
expansion at constant entropy, so meaning with no radiation production.
Thus if inflation started at finite temperature, such as in new
inflation scenarios \cite{ni}, the universe would supercool during the
inflation phase, and if inflation started near zero temperature, such as
in chaotic inflation scenarios \cite{ci}, the universe would remain
there during the inflation phase. Then in the second phase of inflation,
radiation is reintroduced or introduced in a period called reheating. In
this letter we show that for a large number of commonly studied
inflation models, this picture is flawed. We show that in many models
significant radiation production occurs during the inflation period,
which has noticeable observational effects on energy density fluctuations
as well as in cases completely altering the evolution history of the
inflaton.

We examine the model of a scalar inflaton field in interaction
with other scalar $\chi$ and fermion fields through the basic
interaction terms

\begin{equation}
{\cal L}_I = - \frac{1}{2}g^2 \phi^2 \chi^2 - 
g' \phi {\bar \psi_{\chi}} \psi_{\chi} - h \chi {\bar \psi_d}\psi_d ,
\label{lint}
\end{equation}

\noindent
where hereafter $\phi$ is the inflaton field, $\chi$ and
$\psi_{\chi}$ are additional fields to which the inflaton couples, and
$\psi_d$ are light fermions into which the scalar $\chi$-particles can
decay $m_\chi > 2m_{\psi_d}$. Aside from the last term, this is the
typical Lagrangian used in studies of reheating after inflation
\cite{reheato, reheat}, or more to the point, if the inflaton is to eventually
release its energy into the universe, such interactions are to be
expected. Moreover in multifield inflation models, such as hybrid
inflation \cite{hi}, the bosonic interaction plays a basic
role in defining the model.

We can qualitatively divide the study of the interactions (\ref{lint})
into two regimes, weak coupling $g,g' \lesssim 10^{-4}$ and
moderate to large perturbative coupling $10^{-4} \lesssim g,g'
\lesssim 1$. In the weak coupling regime the radiative corrections are
too small to affect the flatness of the the inflaton effective
potential. In the moderate to large perturbative coupling regime, the
radiative corrections from individual terms are significant, so some
mechanism, like supersymmetry, must be invoked to cancel these effects
and maintain an ultraflat inflaton effective potential.
Later in the
paper we will show that in minimal SUSY extensions of the typical
reheating model, decay channels for the $\chi$ or $\psi_{\chi}$
particles are present and the $\psi_d$ field above is simply a
representative example. Since in the moderate to strong perturbative
regime, reheating models will require SUSY for controlling radiative
corrections, Eq. (\ref{lint}) with inclusion of the $\psi_d$ field thus
is a toy model representative of the typical reheating model.

Before proceeding, the regime in which the
radiation energy density during inflation is below the Hubble scale
$\rho_r \lesssim H^4$ will be referred to as isentropic, supercooled, or
cold inflation \cite{oldi,ni,ci,reheato,reheat,hi} and the other regime $\rho_r
> H^4$ will be referred to as nonisentropic or warm inflation \cite{wi}.
As referred to Eq. (\ref{lint}), this letter will focus on the moderate
to large perturbative coupling regime.

In the conventional approach \cite{oldi,ni,ci,reheato,reheat,hi}, for the
effective evolution equation of the inflaton background component,
$\varphi \equiv \langle \phi \rangle$, the assumption has been
that aside from radiative corrections that modify the
$\varphi$-effective potential, $V_{\rm eff}(\varphi)$, this equation is
the same as its classical counterpart. However, we have shown in earlier
works \cite{gr,bgr,br}, that in addition to radiative corrections,
quantum effects also arise in the $\varphi$-effective equation of motion
(EOM) in terms of temporally nonlocal terms, which generically lead to
dissipative effects. Moreover, although SUSY cancels large quantum
effects in the local limit, for the dynamical problem the nonlocal
quantum effects can not be cancelled by SUSY. Our previous considerations
of inflaton dynamics were limited to nonexpanding spacetime. Here we
have extended the calculation to the expanding case (for related earlier
works see \cite{morikawa,idl89}). We also provide in this work possible
physically viable particle physics models, including those
based on SUSY, that realize the dissipative dynamics being considered.
In this letter we will not delve too much into the
(sometimes) evolved technical details of the calculations and
the quantum field theory dynamics, for which we
refer the interested reader to our previous papers
\cite{br} and the most recent
one \cite{br4}.  Here we will focus on the
application and relevance of the calculations to inflation and cosmology
in general. 
{}For
this, in Sec. \ref{sect2} we start by giving the results for the
inflaton effective equation of motion that emerges from
a response theory derivation that 
generalizes to expanding spacetime our previous calculations \cite{br}.
Then in Sec. \ref{sect3} we present physically viable particle physics models that
can realize the typical interactions and decaying mechanism leading to
our results, discussing in particular models based on supersymmetry
where large quantum corrections, that otherwise could spoils the usual flatness
requirements for the inflaton potential, can be kept under control.
A minimal SUSY model is explicitly presented and its properties discussed.
{}Finally the conclusions are given
in Sec. \ref{concl}.

\section{The Equation of Motion and Dissipation in de Sitter Spacetime}
\label{sect2}

Our calculation here is a
generalization to expanding spacetime of our previous results \cite{br}
and whose details are also reported in 
\cite{br4}. 
The starting point is the full equation of motion for the inflaton,
taken as a homogeneous classical background, $\varphi\equiv\varphi(t)$, 
which for the couplings shown in Eq. (\ref{lint}) gives

\begin{eqnarray}
\ddot{\varphi}(t) + 3H(t) {\dot \varphi}(t) +  
\xi {\rm R}(t) \varphi(t) +
\frac{d V(\varphi)}{d\varphi}
+ g^2 \varphi (t) \langle \chi^2 \rangle +
g' \langle {\bar \psi_{\chi}} \psi_{\chi} \rangle =0 \;,
\label{eqphi}
\end{eqnarray}

\noindent
where $H(t)= \dot{a}/a$ is the Hubble parameter, ${\rm R}(t)$ the
curvature scalar, and $\xi$ the gravitational coupling. 
$V(\varphi)$ is the tree level 
potential for the inflaton.  In the calculations that
follow,
we will choose this to be a quartic potential $\lambda \phi^4/4$.
To satisfy density perturbation constraints, it is well
known the self coupling is tiny, $\lambda \sim 10^{-13}$,
which is why in writing Eq. (\ref{eqphi})
we have neglected the insignificant quantum fluctuation
corrections coming from the self-interaction of the inflaton field.
Thus for our demonstration of dissipation and radiation production,
in the $\varphi$ effective EOM it is enough to consider the inflaton 
field as a classical background field interacting with a scalar
$\chi$ and spinor $\psi_{\chi}$ quantum field.
As shown previously \cite{br}, the two quantum
correction terms in Eq. (\ref{eqphi}) lead to terms contributing
both in the linear regime with respect to 
the inflaton amplitude from the Yukawa
interaction due to the fermionic
quantum corrections and in the nonlinear regime 
from the quadratic coupling of $\chi$ to the inflaton.

So far we have not said anything about the role of the $\psi_d$
spinors in Eq. (\ref{lint}) and their contribution in Eq. (\ref{eqphi}).
Their main effect is to dress the propagator for the $\chi$ scalar
and so they enter nonperturbatively in Eq. (\ref{eqphi}) through the
loops of the $\chi$ field. In addition, if we also allow open decay
channels of $\chi$ into $\psi_d,\bar{\psi}_d$, with decay rate
$\Gamma_{\chi \to \psi_d,\bar{\psi}_d}$, the main effect of these
spinors is to provide an effective damping 
in the dressed $\chi$ Green's functions. This will in turn reflect back in the
$\varphi$ effective EOM Eq. (\ref{eqphi}) as an effective 
dissipation for the inflaton field arising from the 
$g^2 \varphi (t) \langle \chi^2 \rangle$ term.
If the inflaton is
also allowed to decay into the $\psi_\chi$ fermions, this can
manifest itself in Eq. (\ref{eqphi}) as a dissipation term due to the
inflaton direct decay and will be important for instance in the linear 
regime for the inflaton amplitude, or more specifically, at the time of
reheating when the inflaton may oscillate around the minimum
of its effective potential. These latter effects that can come from
the fermionic quantum correction in Eq. (\ref{eqphi}) are not the effects 
we are interested
in here; these effects have been well described in several previous studies
of reheating after inflation \cite{reheato,reheat}.  
Here we are concerned instead with the
dissipation identified in Refs. \cite{br,br4}, associated with the 
nonlinear regime for the inflaton field, which can
manifest during inflation; these effects mainly arise from the
nonlocal quantum corrections due to the $\chi$ scalars in Eq. (\ref{eqphi}). 
As shown in our previous references, we can then express Eq. (\ref{eqphi}) 
in terms of a nonlocal effective EOM, relevant in the nonlinear regime,
in the form  

\begin{eqnarray}
{\ddot \varphi}(t) & + & 3H(t) {\dot \varphi}(t) + 
\xi {\rm R}(t) \varphi(t) + \frac{dV_{\rm eff}(\varphi(t))}{d\varphi(t)}
\nonumber \\
& + & 4g^4 \varphi(t) \int_{t_0}^t dt' \varphi(t') {\dot \varphi}(t') K(t,t')
=0,
\label{eom1}
\end{eqnarray}
where \cite{br4}

\begin{eqnarray}
K(t,t') =  \int_{t_0}^{t'} dt'' \int \frac{d^3q}{(2\pi)^3}
\sin \left[2\int_{t''}^t d\tau \omega_{\chi,t}(\tau)\right] 
\frac{\exp\left[-2
\int_{t''}^t d\tau \Gamma_{\chi,t} (\tau)\right]}
{4\omega_{\chi,t}(t) \omega_{\chi,t}(t'')} ,
\label{kernel}
\end{eqnarray}

\begin{eqnarray}
\omega_{\chi,t}(\tau) = \left[{\bf q}^2 a^2(t)/a^2(\tau) 
+ m_{\chi}^2 +
2(6\xi -1)H^2 \right]^{1/2},
\label{omega}
\end{eqnarray}

\noindent
$m_{\chi}=g\varphi \gg m_{\psi_d}$, and $\Gamma_{\chi,t}(\tau) \simeq h^2
m_{\chi}^2/[8\pi \omega_{\chi,t}(\tau)]$ is the decay width of scalars
$\chi$ of comoving momentum ${\bf q}$
into fermions $\psi_d$. The scale factor has been chosen $a(t) = \exp(Ht)$, 
since we  restrict the analysis to
the inflationary, quasi-deSitter regime. $H = \sqrt{8 \pi V_{\rm
eff}/(3m_{\rm Pl}^2})$ is the Hubble parameter,
with $m_{\rm Pl}$ the Planck mass, and 
$V_{\rm eff}(\varphi(t))$
is the renormalized effective potential for the inflaton, after the
local quantum corrections to $V(\varphi)$ have been taken into account.

The kernel $K(t,t')$, Eq. (\ref{kernel}), is obtained by a 
response theory approximation, which takes advantage of the slow
dynamics of the inflaton field so that the dynamics happen effectively
in the adiabatic regime.  This is equivalent to the
closed time path formalism at leading nontrivial order \cite{gr,bgr}, which
treats the effect of the field $\chi$ on the evolution of $\varphi(t)$.
The mode functions for the $\chi$-field that enter in the Green's functions
appearing in the response theory expansion of the $\langle \chi^2 \rangle$ term
in Eq. (\ref{eqphi}) (see e.g. Ref. \cite{br}) are computed from 
a WKB approximation, which treats expansion effects and the time
variation of the background inflaton field.
These mode functions are then used in the loop calculations that
determine the effect of the $\chi$ quantum corrections to
the background inflaton EOM.
This WKB approximation is generally valid 
in the adiabatic regime 
$d{\omega}_\chi/d\tau \ll \omega^2_{\chi}$.
In our calculation this adiabatic regime emerges since for the parameter
values we will be studying
the $\varphi$ motion is overdamped and since the $\chi$ mass is
heavy $m_\chi \gg H$. In fact, these two conditions, the slow $\varphi$ dynamics
and the heavy $\chi$ mass are enough to assure the validity of a WKB 
approximation for the $\chi$ modes. In addition, the condition 
$m_{\chi} \gg H$ implies that
curvature effects in the $\chi$ field quantum corrections to the
background inflaton field are subleading, with dominant term being the
Minkowski like corrections. This will be explicitly 
observed below when we numerically
compare several results based on Eq. (\ref{eom1}). 
Note, in the limit $H \rightarrow 0$, $a \rightarrow
constant$, Eq. (\ref{kernel}) agrees with the corresponding kernel for
nonexpanding spacetime in \cite{br}.
The physical origin of the nonlocal
(dissipative) term in Eq. (\ref{eom1}) is as follows. The evolving
background field $\varphi$ changes the mass of the $\chi$ boson which
results in the mixing of its positive and negative frequency modes. This
in turn leads to coherent production of $\chi$ particles which then
decohere through decay into the lighter $\psi_d$-fermions.

When the motion of $\varphi$ is slow, 
an adiabatic-Markovian approximation can be applied that converts Eq.
(\ref{eom1}) to one that is completely local in time, albeit with time
derivative terms (for Minkowski spacetime 
details of this approximation are in \cite{br}, while for the expanding
spacetime case this is derived in \cite{br4}). 
The Markovian approximation amounts to substituting $t' \rightarrow t$ in
the arguments of the $\varphi$-fields in the nonlocal term 
in Eq. (\ref{eom1}).
The Markovian approximation then requires self-consistently
that all macroscopic motion is
slow on the scale of microscopic motion, thus
${\dot \varphi}/{\varphi},H < \Gamma_{\chi}$.
Moreover when $H < m_{\chi}$,
the kernel $K(t,t')$ is well approximated by the
nonexpanding limit $H \rightarrow 0$. 
The validity of all these approximations will be examined below.
Combining these
approximations, the effective EOM Eq. (\ref{eom1}) becomes

\begin{equation}
{\ddot \varphi} + [3H+\Upsilon(\varphi)] {\dot \varphi} +
\xi {\rm R} \varphi + \frac{dV_{\rm eff}(\varphi)}{d\varphi} = 0,
\label{amapprox}
\end{equation}
where, by defining $\alpha_\chi = h^2 /(8 \pi )$,
                                                                                
\begin{equation}
\Upsilon(\varphi) = \frac{\sqrt{2} g^4 \alpha_\chi \varphi^2 }
{64\pi m_\chi \sqrt{1 + \alpha^2_{\chi}}
\sqrt{\sqrt{1 + \alpha^2_{\chi}}+1}} .
\end{equation}

\begin{figure}[ht]
\vspace{1cm}
\epsfysize=5.95cm
{\centerline{\epsfbox{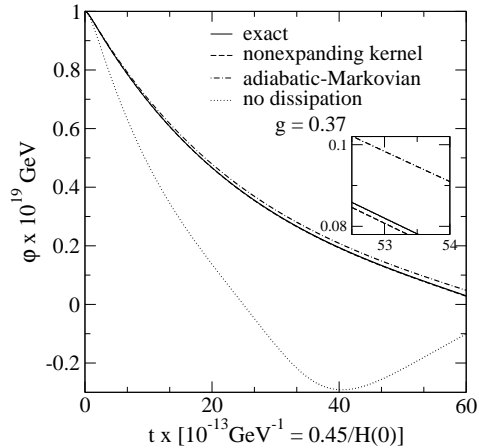}}}
\caption{Evolution of $\varphi(t)$ for $\lambda=10^{-13}$,
$g=h=0.37$, $\xi=0$,
$\varphi(0) = m_{\rm Pl}$, ${\dot \varphi}(0)=0$.}
\label{fig1}
\end{figure}

{}Fig. \ref{fig1} compares the various approximations for a
representative case where $g=h=0.37$ and the inflaton potential is that
for chaotic inflation $V_{\rm eff}(\varphi) = \lambda \varphi^4/4$
with $\lambda = 10^{-13}$ \cite{ci}. In {}Fig \ref{fig1} evolution has
been examined at the final stages of chaotic inflation where we start
with $\varphi(t_0=0) = m_{\rm Pl}$. The solid line is the exact result based on
numerically solving Eq. (\ref{eom1}). Plotted alongside this, although
almost indiscernible, is the same solution expect using the nonexpanding
spacetime kernel (dashed line), obtained by setting $H \rightarrow 0$,
$a \rightarrow constant$ in Eq. (\ref{kernel}), and the solution based
on the adiabatic-Markovian approximation of Eq. (\ref{amapprox})
(dot-dashed line) for the same parameter set. As seen, the expanding and
nonexpanding cases differ by very little and the adiabatic-Markovian
approximation is in good agreement with the exact solution. This
confirms simplifying approximations claimed in \cite{bgr,br,morikawa}
but up to now had not been numerically verified.

More interestingly, and the first major result of this letter, the
dotted line in {}Fig. \ref{fig1} is the solution that would be found by
the conventional approach in which the nonlocal term in Eq. (\ref{eom1})
is ignored.  The conventional approach \cite{ni,ci,reheato,reheat,hi}
expects the inflaton to start oscillating, which is the precursor to
entering various stages of pre/re-heating. However with account for
dissipative effects, this never happens for our example in {}Fig.
1, since the inflaton remains overdamped till the end when it settles at
its minima at $\varphi=0$. Moreover, throughout inflation, and not just
near the end, the inflaton dissipates energy, which yields a
radiation component of magnitude

\begin{equation}
\rho_r \approx \Upsilon {\dot \varphi}^2/(4H) \;.
\label{rad}
\end{equation}

{}For the case in {}Fig. \ref{fig1}, the overdamped regime for the
inflaton persists until $g \lesssim 0.35$, below which its evolution at
the end of inflation has oscillatory features similar to conventional
expectations. However the inflaton is still dissipating radiation at the
level Eq. (\ref{rad}) all throughout the inflation period. The radiation
energy density produced through this dissipative mechanism is greater
than the Hubble scale, $\rho_r > H^4$ and $\Gamma_{\chi} > H$, if for
example $g=h$ and $g > 10^{-2}$ or, as another example, if $h=0.1$ and
$g > 10^{-3}$. If this radiation thermalized, which is expected for
$\Gamma_{\chi} > H$, then $T \approx \rho_r^{1/4} > H$ and it is known
that the primordial density perturbations, $\delta \rho /\rho$, produced
by the inflaton are altered from their zero temperature value
\cite{wi,Moss:wn,Berera:1995wh,Berera:1999ws}. In particular in the
standard relation $\delta \rho/\rho \sim H \delta \varphi/{\dot
\varphi}$, for $T < H$, $\delta \varphi^2 \sim H^2$ is the cold
inflation result \cite{ni,ci,hi}, for $T > H$ and $\Upsilon < 3H$,
$\delta \varphi^2 \sim HT$ \cite{Moss:wn,Berera:1995wh}, and for
$T,\Upsilon/3 > H$, $\delta \varphi^2 \sim \sqrt{H\Upsilon}T$
\cite{Berera:1999ws}.
Applying these relations to obtain the spectral index $n_s$
for the $\lambda \phi^4/4$ model, for cold inflation the known
value is $1-n_s = 3/N_e$ \cite{Liddle:1992wi}, whereas in contrast for
strong dissipative warm inflation ($\Upsilon > 3H$),  for the same
normalization we find $1-n_s = 3/(2N_e)$, where $N_e$ is the number
of e-folds of inflation.
For example for $N_e = 60$ 
the discrepancy in $n_s$ between the warm and cold
inflation cases is about 3\%, thus possibly resolvable
with current generation high-precision satellite
experiments, WMAP and Planck.
Moreover cold inflation for any e-folding $N_e > 1$ has
the well known $\eta$ and large $\varphi$-amplitude problems,
since observationally consistent slow-roll solutions 
require $m_{\phi} < H$ and $\varphi > m_{\rm Pl}$ 
respectively \cite{Liddle:1992wi}.  On the other hand,
for warm inflation these problems are nonexistent,
since we find for example under the observational
constraints $\delta \rho/\rho = 10^{-5}$
and $N_e=60$, 
$m^2_{\phi} \approx 190H^2 > H^2$ and
$\varphi \approx 0.087m_{\rm Pl} < m_{\rm Pl}$.
We thus arrive at our second major result.  
In multifield inflation models, or in fact any
inflation model once reheating interactions are accounted for,
there are parameter regimes feasible to inflation,
that have never been properly treated since 
the nonlocal term in Eq. (\ref{eom1}) is neglected.
Once this term is retained,
it is seen that up to fairly weak coupling,
adequate radiation is produced during inflation to alter
density perturbations.
Although this conclusion is
based on situations where thermalization is assumed,
one can infer the same qualitative 
outcome once $\rho_r > H^4$,
irrespective of the statistical state.

In regimes where Eq. (\ref{eom1}) produces significant radiation during
inflation, models such as in \cite{reheat,hi,embed} have to be examined
on a case by case basis to determine the effect on density
perturbations. Dissipative effects themselves can produce a rich variety
of spectra ranging between red and blue
\cite{Moss:wn,denpert,Hall:2003zp}.  A dramatic example is in
\cite{Hall:2003zp} which treats new inflation type symmetry breaking
potentials.  In cold inflation, such models yield relatively
featureless red spectra, whereas with account for dissipative effects, the
spectra can be altered between red and blue producing, as an example,
the running blue to red spectra suggested by WMAP/2df/Sloan
\cite{wmap}.

\section{SUSY Model Implementations and Particle Decaying Channels}
\label{sect3}

The above results rely on the ability of the $\chi$-particles to decay,
which in our toy model Eq. (\ref{lint}) is fulfilled by the
$\psi_d$-fermions. 
Based on Eq. (\ref{kernel}), this dissipative
mechanism is not specific to what the $\chi$ particles decay into, but
simply requires a nonzero $\Gamma_{\chi} > H$. Thus the typical
possibilities for a scalar field are either decaying into fermions, as
in our toy model, or into gauge bosons. Also, in models where $\chi$
develops a nonzero vacuum expectation value, 
such as hybrid inflation \cite{hi}, the $\phi -
\chi$ interaction generates the term $g \langle \chi \rangle \chi
\phi^2$, which implies a direct $\chi \rightarrow \phi \phi$ decay
channel. Independent of inflation issues, particle physics models are
adept with interaction configurations similar to Eq. (\ref{lint}). 
For
example, the simplest implementation of the Higgs mechanism in the
Standard Model has the background Higgs field coupled to W and Z bosons,
thereby generating their masses, and these bosons then are coupled to
light fermions through the well known charged and neutral current
interactions. 
In the context of inflation, conventional studies
typically only present the inflaton sector itself, since the tacit
assumption in these studies has been that interactions are unimportant
to the inflation phase. The lesson learned by this work is interactions
can play a vital role within the inflation phase, and so care must be
taken to understand how the inflaton sector is embedded within the full
theory. The few studies that have attempted to do this \cite{embed}
contain interaction structures similar to Eq. (\ref{lint}).

Whether in our model Eq. (\ref{lint}), reheating models \cite{reheat},
or any other inflation model which has a $\phi - \chi$ coupling in the
moderate to large perturbative regime \cite{hi}, sizable
corrections are induced to the $\varphi$-effective potential 
and so SUSY must be
invoked. A minimal SUSY extension that incorporates the $\phi-\chi$
coupling would be

\begin{equation}
W= \sqrt{\lambda} \Phi^3 + g \Phi X^2 + f X^3 + m X^2,
\label{susymodel}
\end{equation}

\noindent
where $\Phi = \phi + \psi \theta + \theta^2 F$ and $X = \chi+
\theta \psi_{\chi} + \theta^2 F_{\chi}$ are chiral superfields. In the
above model, there would be no additional fermion to associate with
$\psi_d$ from our toy model Eq. (\ref{lint}). However the $\chi$-field
has a decay channel via a $\psi_{\chi}$ triangle-loop into two light
inflaton bosons $\phi$. {}For this case, everything in Eqs.
(\ref{eom1})-(\ref{omega}) is unaltered except the decay channel is
different with now $\Gamma_{\chi} \sim (fg^2)^2 m_{\chi}$. Thus there
are additional factors of coupling constants, but in the large
perturbative regime, the evolution of $\varphi$ should be similar to the
solid line in {}Fig. \ref{fig1}.
Similarly the $\psi_{\chi}$-fermion can decay into $\phi$
and $\psi$, mediated through a triangle loop comprised of heavy $\chi$
and $\psi_{\chi}$ particles. 

Other weaker channels
also occur. {}For example,
the $\chi$ and $\psi_{\chi}$ SUSY partners can have
masses differing by $\Delta m \gtrsim m_{\phi}$, 
without harming the one-loop cancellations in the $\varphi$ effective
potential.
Thus in the model Eq.(\ref{susymodel}), a $\chi$ particle could
decay into two fermions, like in our toy
model Eq. (\ref{lint}), with one being the fermion partner $\psi_{\chi}$
and the other an inflatino $\psi$.  This channel is phase space
suppressed since $m_{\chi} - m_{\psi_{\chi}} \sim m_{\phi}$, which
leads to $\Gamma_{\chi} \sim g^2m^2_{\phi}/m_{\chi}$.  
Based on Eqs. (\ref{amapprox}) and (\ref{rad}),
this channel leads to $T > H$ for 
$g \gtrsim 0.3$.  Weaker channels
such as these need not dominate
radiation production but also can serve other
particle physics functions such as  baryogenesis \cite{wi,by}
or dark matter production.

Up to now very little has been said about the role of the
$\psi_{\chi}$-fermion in our toy model.  Here its only purpose is to supply a
one loop contribution that can suppress the large contributions from the
$\chi$ sector (then four $\chi$ fields are required in Eq.
(\ref{lint})), thus mimic SUSY \cite{bgr,br}.
For de Sitter spacetime, these results are modified due to $\xi$-dependent
mass corrections and the $k_0$ integration in this geometry.
Both modifications add corrections from the Minkowski spacetime
effective potential by terms 
$\sim O(\ll 1) g^2 H^2 \varphi^2  < m^2_{\phi} \varphi^2$ 
\cite{Vilenkin:sg}
and so can be neglected. Additional analysis of both quantum
and thermal corrections
for SUSY models like 
Eq. (\ref{susymodel}) has been studied independently by the authors of
Ref. \cite{moss2}.  They have also concluded
that quantum corrections can be kept under control and in
addition they find the same holds for thermal corrections.

As briefly discussed below Eq. (\ref{rad}), for warm inflation 
in the strong dissipative regime, an important result is the
avoidance of the $\eta$-problem.  In cold inflation models,
the $\eta$-problem is that 
slow-roll inflation requires $m_{\phi} < H$ whereas particle
physics effects and the coupling to the background metric often prohibit
this from happening. In particular in SUSY models, since they imply
supergravity, it means higher dimensional operators inherently enter in
forms such as $a_n \phi^n/m_p^{n-4}$ or $V \phi^2/m_{p}^2$, which at
either large or small field amplitude can cause the $\eta$-problem.
However for warm inflation in the strong dissipative regime, slow-roll
and density perturbation requirements can be satisfied for $m_{\phi} >
H$ and even $m_{\phi} \gg H$
\cite{wi,Berera:1999ws,denpert,Hall:2003zp}, thus intrinsically there is
no $\eta$-problem in any model.

\section{Conclusions}
\label{concl}

It is now opportune to give a few remarks on the applications and
limitations of our calculation. 
Eqs. (\ref{eom1}) and (\ref{kernel}) are valid
for any decay channel for the $\chi$-boson, thus any
$\Gamma_{\chi}$, including the case of no decay channel,
$\Gamma_{\chi} =0$. 
If $\Gamma_{\chi} > H$, then the approximation Eq. (\ref{amapprox}) also
applies.
In the regime,
$\Gamma_{\chi} < H$, within the characteristic oscillation period
of the free inflaton field $ \sim 1/m_{\phi}$, $K(t,t')$ oscillates
considerably, since $m_{\chi} \gg m_{\phi},H$ and without attenuation,
since $\Gamma_{\chi} < H$. Numerically this is a more difficult regime
for studying the net effect of the nonlocal term.
Another point to note is that Eq. (\ref{eom1})
only has the temperature independent contribution to dissipation and so
only provides a lower bound. {}For $\Gamma_{\chi} > H$ thermalization can
in principle be
justified and when the temperature is bigger than either the inflaton
mass, $T> m_{\phi}$, or the decay product mass, $T> m_{\psi_d}$ in our toy
model, dissipative effects are enhanced \cite{bgr,gr,RF}. 
In further work, all these points need to be investigated.
Although this work focused on inflaton dissipative effects induced by
a $\chi$-boson field, similar considerations can be applied for a
$\psi_{\chi}$ fermion. This implies that fermionic and
bosonic pre/reheating may be equally affected by our results. In any event,
for moderate to large perturbative coupling, SUSY forces both types of
interactions to be present side-by-side.

Inflation and reheating phases have been in the past treated as almost
detached problems, as so has the quantum-to-classical
and the initial
condition problems and so on. On the other hand, the warm inflation
picture shows, here and elsewhere \cite{wi,denpert,Hall:2003zp},
that inflation can be treated
as a single consolidated scenario.
As one outcome, various problems,
namely $\eta$,
graceful exit, quantum-to-classical transition, large inflaton
amplitude, and aspects of initial conditions \cite{bg,ror}, can be
nonexistent in warm inflation.

\acknowledgments

We thank Robert Brandenberger, Tom Kephart and Ian Moss 
for helpful discussions.
A.B. is funded by UK PPARC  
and R.O.R. is partially funded by CNPq-Brazil.

\end{document}